# *Solaris*: A Focused Solar Polar Discovery-class Mission to achieve the Highest Priority Heliophysics Science Now

**Primary Author:** Donald M. Hassler[1], Southwest Research Institute, Boulder, CO

**Co-Authors:** Sarah E Gibson[2], Jeffrey S Newmark[3], Nicholas A. Featherstone[1], Lisa Upton[1], Nicholeen M Viall[3], J Todd Hoeksema[4], Frederic Auchere[5], Aaron Birch[6], Doug Braun[7], Paul Charbonneau[8], Robin Colannino[9], Craig DeForest[1], Mausumi Dikpati[2], Cooper Downs[10], Nicole Duncan[11], Heather Alison Elliott[1], Yuhong Fan[2], Silvano Fineschi[12], Laurent Gizon[6], Sanjay Gosain[13], Louise Harra[14], Brad Hindman[15], David Berghmans[16], Susan T Lepri[17], Jon Linker[10], Mark B. Moldwin[17], Andres Munoz-Jaramillo[1], Dibyendu Nandy[18], Yeimy Rivera[19], Jesper Schou[6], Justyna Sokol[1], Barbara Thompson[3], Marco Velli[20], Thomas N. Woods[14], Junwei Zhao[4]

**Institutions:** [1]SwRI, [2]NCAR, [3]GSFC, [4]Stanford, [5]IAS, [6]MPS, [7]NWRA, [8]U. Montreal, [9]NRL, [10]PSI, [11]Ball, [12]INAF, [13]NSO, [14]PMOD, [15]CU, [16]ROB, [17]U. Mich., [18]IISER, [19]CfA, [20]UCLA

## Synopsis

*Solaris* is a transformative Solar Polar Discovery-class mission concept to address crucial outstanding questions that can only be answered from a polar vantage. *Solaris* will image the Sun's poles from 75° latitude, providing new insight into the workings of the solar dynamo and the solar cycle, which are at the foundation of our understanding of space weather and space climate. Solaris will also provide enabling observations for improved space weather research, modeling and prediction, revealing a unique, new view of the corona, coronal dynamics and CME eruptions from above.

The *Solaris* mission design is an ***expanded/enhanced*** version of the mission design from the MIDEX Phase A study of the same name. Solaris includes ***both remote sensing and in-situ instruments*** that are essential to address fundamental questions that can only be answered from a polar perspective. ***Solaris' ~10 yr mission covers the solar cycle*** and ***achieves multiple solar polar passes*** using a simple, ballistic trajectory. ***Solaris is ready to go now***, moving Heliophysics forward at a critical time when models and computing power are ideally suited to capitalize on new polar observations. *Solaris* ***uses existing technology*** and can stand alone as a ***single spacecraft mission*** targeting ***questions that cannot wait***, or act as the first element of a disaggregated constellation mission. Solaris is a ***prerequisite to inform the design*** (both science observations and technical mission) ***of potential future flagship missions*** before investing $Billions.

### Solaris – Innovative, Game-Changing Mission in a Nutshell

*Solaris* enables the highest priority Heliophysics science to be done now. Key elements include:

- ***New insight into the Solar Dynamo*** – the foundation to our understanding of the solar cycle.
- ***Enabling Observations for Space Weather Research*** capitalizing on *Solaris'* "sunny-side up" view of earth-directed transients.
- ***Highly Targeted Payload*** to address focused science questions.
- ***PI-led Mission Paradigm*** based on Discovery mission concept, enabling rapid implementation while PSP and Solar Orbiter are still operational.
- ***Heritage Spacecraft Design & Subsystems*** leveraging other existing deep space missions.
- ***Simple Mission Design -*** ballistic trajectory requires minimal contact during cruise & synoptic observations during encounter phases.
- ***Minimalist Approach to Operations*** – Phase E has two components: 1) Quiescent Cruise with minimal staffing, 2) Remote Sensing encounter when *Solaris* is >55° helio-latitude, with simple, synoptic observations planned in advance; no campaigns.

# 1 Executive Summary

*Solaris* is a focused Solar Polar Discovery-class mission concept to address crucial outstanding questions that can only be answered from a polar vantage. *Solaris* images the Sun's poles from 75° latitude, providing new insight into the workings of the solar dynamo and the solar cycle. *Solaris* also provides enabling observations for improved space weather modeling and prediction and reveals a unique, new view of coronal dynamics and coronal mass ejections (CMEs) from above.

The *Solaris* mission design is an ***expanded/enhanced version*** of the mission design from the MIDEX Phase A Study by the same name and includes ***both remote sensing and in-situ instruments***. ***Solaris' ~10 yr mission covers the solar cycle*** and uses a ballistic trajectory and a Jupiter Gravity Assist (JGA) to fly over both poles of the Sun at ≥75° inclination to its equator, and *multiple Venus Gravity Assists (VGAs)* to reduce the aphelion and circularize the orbit, resulting in a ~3 yr orbital period that ***achieves multiple solar polar passes*** with multi-month continuous observations at different times throughout the solar cycle ***(Fig. 1)***.

*Solaris* has been decades in the waiting, originally proposed as part of the International Solar Polar Mission (ISPM) in the 1970s (remote-sensing "sister spacecraft" to Ulysses), and more recently as the Solar Polar Imager (SPI) in the 2013 Decadal Survey. *Solaris* is timely and urgent and ready to go now with existing technology, providing an opportunity for collaborative, multi-viewpoint observations and science with Parker Solar Probe (PSP), Solar Orbiter (SO), Space Weather Follow On (SWFO) at L-1 and the European Vigil mission at L-5.

*Solaris* addresses big, fundamental questions that are still unanswered. How does the solar dynamo generate cycling magnetic fields? How does it drive solar activity and shape the heliosphere over the solar cycle? ***These are the some of the highest priority questions in Heliophysics today.*** We have theories and models, each with their own mechanisms for sustaining the dynamo, but we don't really know which mechanisms are most important for the Sun. ***We don't have the measurements we need*** to distinguish and constrain these models. These measurements, and the answers to these questions, can ***only be answered*** by observing our Sun from a new perspective, ***from over the Sun's poles***, obtaining vital clues to the mechanisms that drive the solar dynamo, and in turn, the solar cycle.

*Solaris* also ***advances Space Weather Research***, a high priority for NASA and the nation, through the first polar magnetograms over multiple solar rotations and simultaneous, 360° longitudinal views of coronal structure, variability, and CMEs. Complete longitudinal coverage enables monitoring of evolving structures and improves modeling of Earth-intersecting transients.

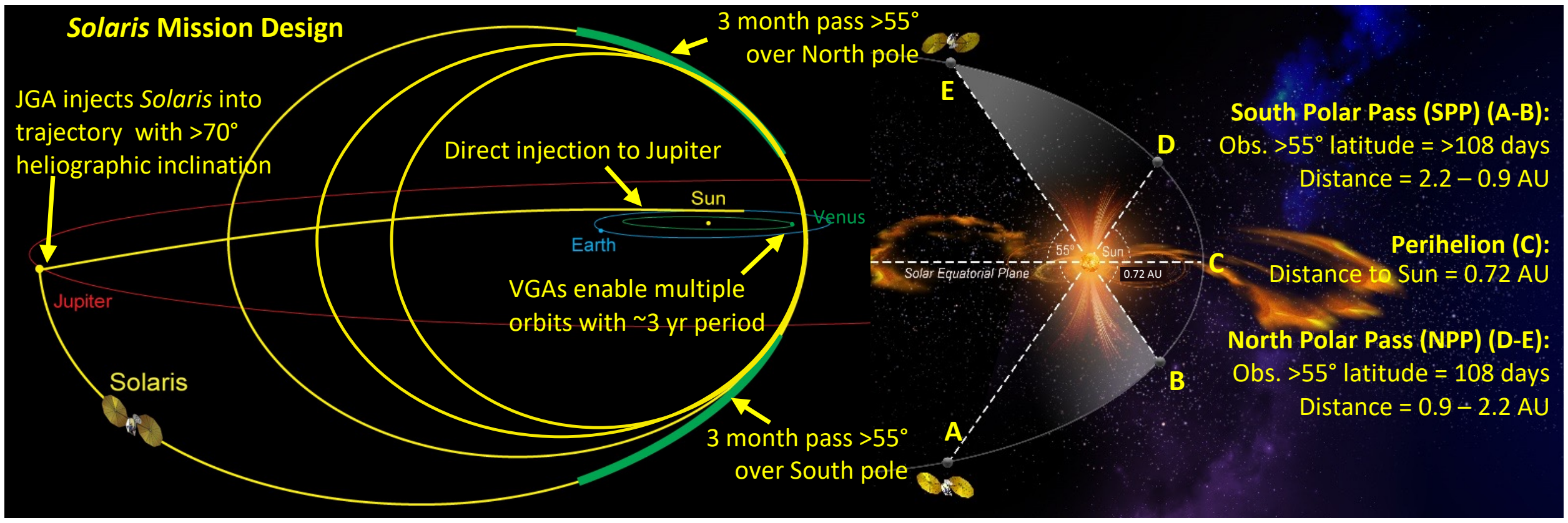

***Figure 1***. ***Solaris observes the never-before-seen poles of the Sun*** *with multi-month polar passes (>55°) over the solar cycle. Solaris' trajectory uses a JGA to escape the ecliptic plane and multiple VGAs to reduce the aphelion and circularize the orbit to a ~3 yr period.*

## 2 Science Objectives

The science enabled by *Solaris* is transformative, filling known gaps in our understanding, and is divided into three objectives.

### 2.1 Solaris Science Objective 1

***To understand how polar magnetic fields and flows reveal the Sun's global dynamics and the mechanisms that underlie the solar dynamo, which ultimately shape the solar activity cycle.*** *How does solar rotation influence convection to determine the Sun's dynamical regime? What is the strength & distribution of the polar magnetic fields & flows that drive solar dynamo poloidal flux generation & transport?*

*Solaris' **Objective 1*** is about the ***dynamo***, which is a ***universal physical process*** fundamental not just to Heliophysics, but to planetary science, earth science and stellar astrophysics (Charbonneau, 2020). The ***missing link*** is polar observation — as recognized within Goal 1 in the 2013 Heliophysics Decadal Survey report: *"The deep, ponderous flows that carry patterns of magnetic flux to the poles regulate the seeding of the deep-seated dynamo that generates subsequent solar cycles…the missing information will come, in part, from measurements of the hard-to-view solar poles" (Decadal Survey 2013).*

***So why do we need to go to the poles to address this Objective?*** Because the solar dynamo is a dance between the Sun's magnetic field and three intrinsically coupled flows (differential rotation, convection and meridional circulation) — all of which are unknown at the poles (Fig. 2).

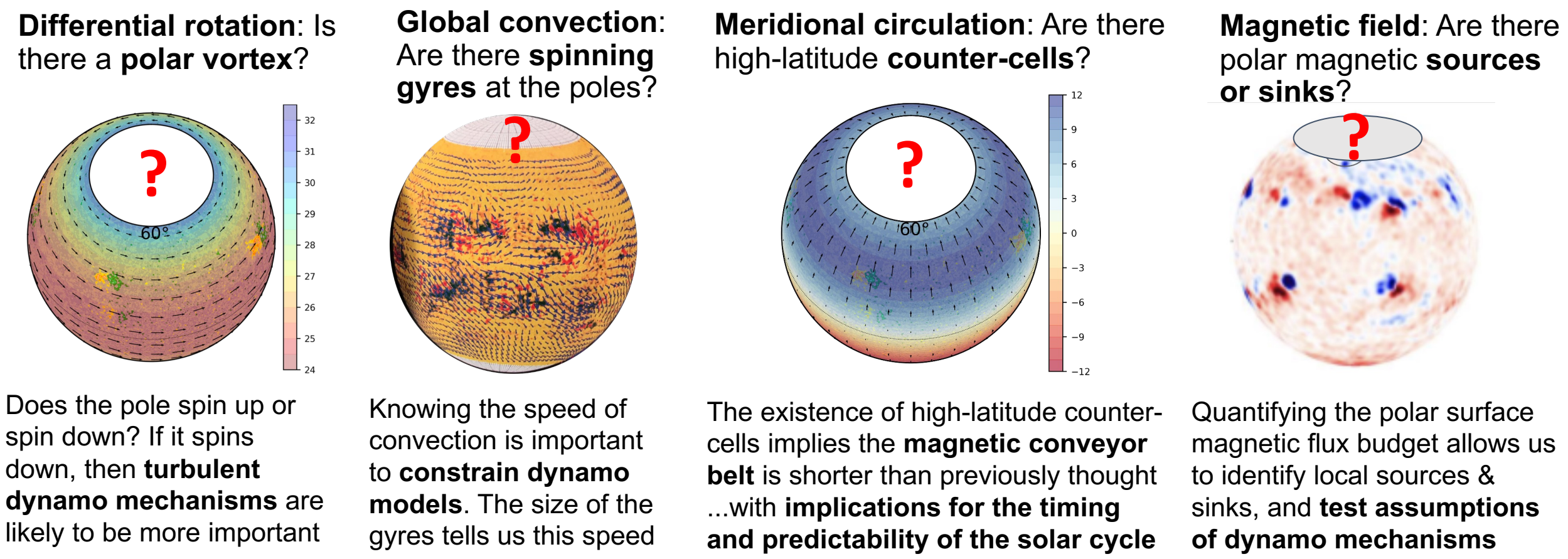

***Figure 2. Why Solaris needs to go to the poles.*** *Solaris' observations of the magnetic field and three intrinsically coupled flows at the Sun's poles are the key to choosing between solar dynamo mechanisms.*

***How does Solaris achieve its Objective 1 in three solar rotations/pole per solar pass?*** *Solaris* achieves closure on this objective through a sustained view of *all polar longitudes simultaneously*, providing a continuous view of high latitudes for >100 days, sufficient to resolve the fields, flows and structures necessary to constrain the mechanisms that form the engine of the solar dynamo and ultimately shape the solar cycle. Moreover, *Solaris* will observe the flows and fields at the heart of solar dynamo mechanisms as a function of time in the solar cycle, providing clues to understanding the variations in solar cycle duration and strength. By measuring near-surface convective flows in the Sun's polar regions, *Solaris* probes the nature of deep convection without having to directly measure the deep convection itself. In all cases, simulations and forward modeling guide *Solaris'* science strategy (Fig. 3; See white paper by Featherstone et al. (2022)).

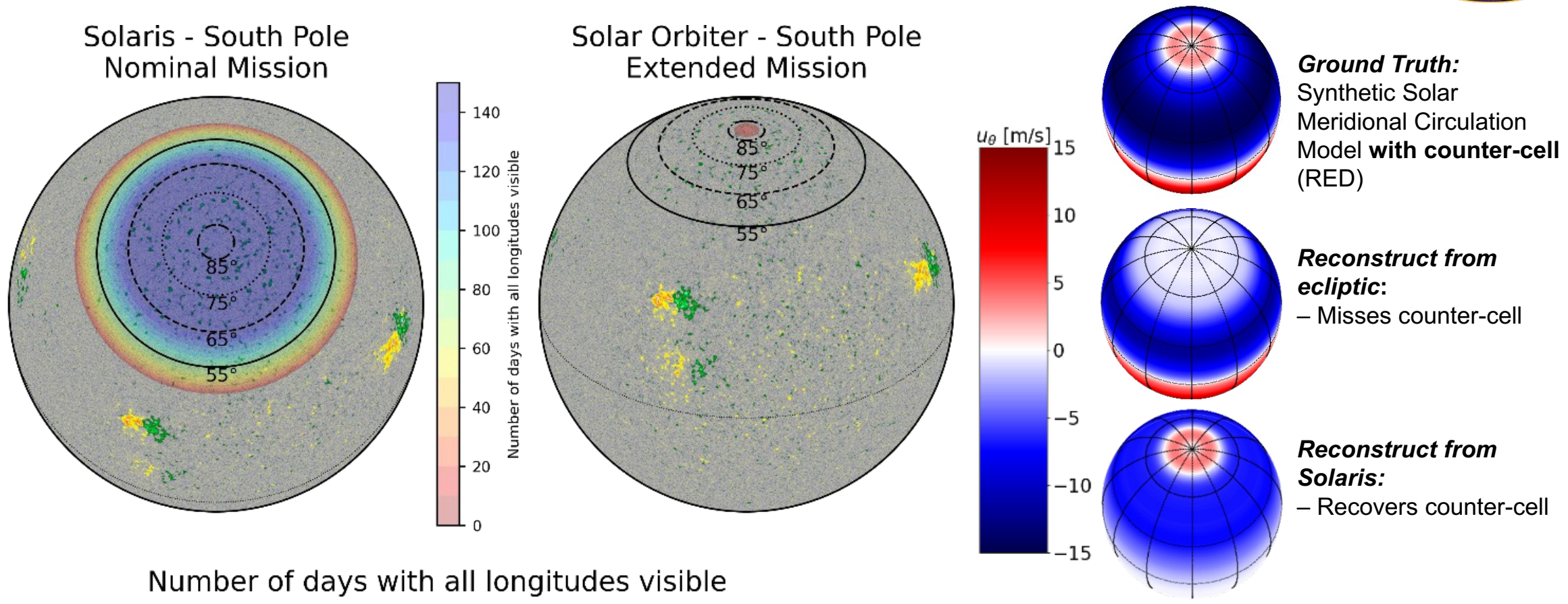

***Figure 3. Sustained high latitude observations are essential to achieving Solaris' objectives. (Left)*** *Foreshortening and other problems limit magnetic field & flow measurements to within 60° of observed solar disk center. The colors show how many days observations in this region are obtained for all 360° of longitude for Solaris (>100 days) and Solar Orbiter (<10 days; even in its extended mission it does not have a sustained view over the top of the polar cap.* ***(Right) Forward modeling shows the importance of sustained, high latitude observations*** *to observe high latitude meridional counter-cells. Solaris recovers polar flows that Solar Orbiter and ecliptic measurements cannot, as demonstrated through a helioseismic inversion of 72 days of synthetic forward-modeled near-surface meridional circulation observations (Gizon et al. 2020; see also white paper by Baldner et al. (2022)).*

***How does Solaris distinguish between the Sun's dynamo mechanisms?*** Different dynamo mechanisms may operate independently or together: the unknown ratio of their relative roles makes dynamo modeling currently unconstrained. Quantifying the degree to which rotation influences convection identifies where the Sun lies on the spectrum of possible stellar dynamical regimes (Fig. 5), and pins down the role of turbulent mechanisms in the solar dynamo. Quantifying the polar magnetic flux budget identifies how the poles are affected by local magnetic sources or sinks, testing assumptions of the Babcock-Leighton dynamo mechanism (see white papers by Upton et al. (2022), and Nandy et al. (2022). ***Solaris explores the relative roles of dynamo mechanisms, and how they may vary over the course of a solar cycle.***

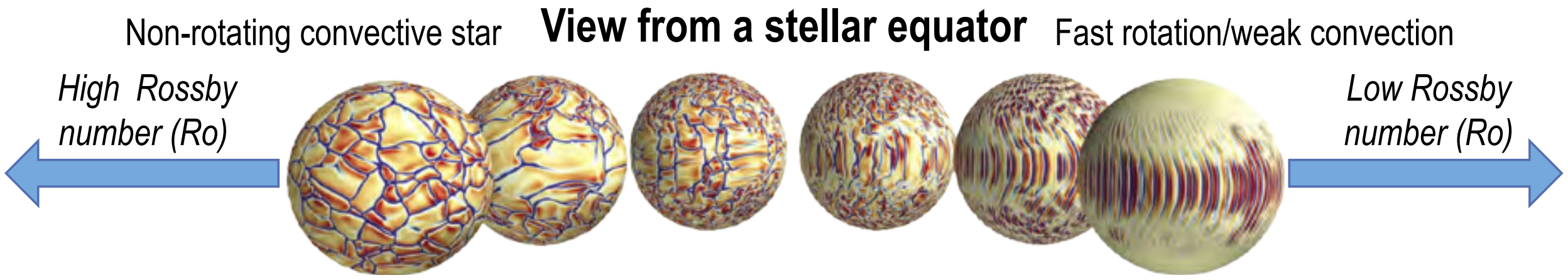

***Fig. 4. Solaris' polar flow measurements pin down the Sun's location on a spectrum of stellar dynamic regimes*** *as seen in convective simulations (Featherstone et al., 2016). The delicate balance between buoyancy and rotation determines the dynamical regime in which a star operates, typically captured by the Rossby number (Ro). Establishing the dynamical regime of the Sun has been difficult. Global simulations shown above allow exploration of (a) high and (b) low Ro regimes, representing (a) weak and (b) strong rotational constraint on convective flow.* ***Solaris' high-latitude vantage allows measurement of convective amplitude and structure, constraining Ro and with it the Sun's dynamical regime.***

## 2.2 Solaris Science Objective 2

***To determine how high-latitude coronal magnetic fields connect the Sun and heliosphere throughout the Solar Cycle.*** *How does the strength, distribution & inclination of open magnetic flux in polar coronal holes compare with global heliospheric magnetic flux? How does the buildup of magnetic energy in closed-field high-latitude structures determine storage and release in coronal eruptions?*

***Solaris' Objective 2*** focuses on the ***interface between the Sun and the heliosphere***, in both magnetically open and closed structures over rotational and solar cycle time scales. Studies of polar coronal holes have shown that the amount of open flux estimated in the corona is systematically lower by a factor of two or more than the amount of magnetic flux measured in situ in the heliosphere (Linker *et al*., 2017, 2021; white paper by Linker et al. (2022)). Without *Solaris*' direct observations of polar magnetic fields, models must guess about the inner boundary of the heliosphere. *Solaris* provides the first global polar magnetic flux observations together with *in-situ* vector magnetic field measurements, enabling direct comparison of spacecraft-observed and coronal-hole-derived estimates of polar magnetic flux. This powerful combination allows ***determination of the solar wind magnetic flux budget and its evolutionary behavior.***

*Ulysses* heavy ion composition measurements revealed that the global heliosphere varied based on solar magnetic structures and solar wind speed, but also provided evidence that processes at work in polar coronal holes were fundamentally different than those that produce the slow solar wind at low latitudes (*e.g.* McComas et al. 2008, Stakhiv *et al.,* 2015 and references therein). *Ulysses* polar observations also showed how turbulence in the polar regions at solar minimum was dominated by outwardly propagating Alfvénic fluctuations, characterized by the presence of a few local magnetic field reversals, or switchbacks. PSP in the inner heliosphere observed the ubiquitous presence of switchbacks embedded in Alfvénic turbulence coming in patches associated with supergranulation and composition changes in-situ (Bale et al, 2021, Fargette et al. 2021). ***Solaris' long period observations at high latitude will reveal*** whether the ***switchback patches relax into micro-streams*** and whether and how they may be ***associated with coronal jets, jet-lets and or emanate from polar coronal holes***.

*Solaris* also studies closed magnetic field regions, in particular polar crown filaments. These regions are important because they are source regions of coronal mass ejections, but operate on significantly larger and longer spatial and temporal scales than CMEs from lower latitude active regions. ***Polar crown filaments are thus the ideal laboratory for studying the build-up of magnetic energy in the corona — in slow motion.***

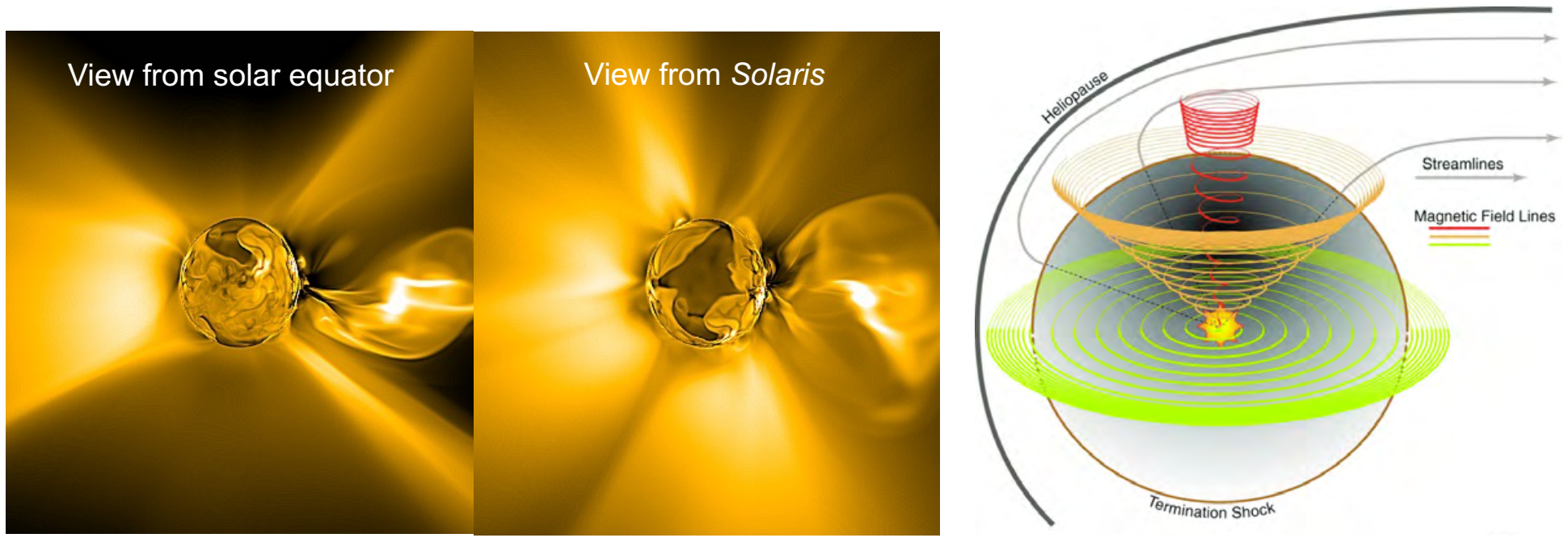

***Figure 5. Solaris has out-of-ecliptic view on solar transients and resolves inconsistencies between global solar wind model and theory***. *(Left) Comparison of CME viewed from the ecliptic plane to CME viewed from Solaris'* polar vantage *(Right) Parker Spiral as function of latitude (Suess, 1999).*

### 2.3 Solaris Science Objective 3

***To determine the role of transient dynamics in structuring the solar wind throughout the Solar Cycle.*** *How does small-scale ecliptic plane variability drive large-scale longitudinal magnetic geometry, and what are the implications for solar wind formation? What is the global ecliptic coronal/solar wind reaction to the early phase/passage of large transients, and what are the implications for their evolution in the solar wind?*

*Solaris'* ***Objective 3*** focuses on ***dynamic connections between the Sun and Heliosphere****.* Deep-field imaging of the mid to outer corona (DeForest et al. 2018) reveals faint moving structures and inhomogeneities at all latitudes and solar wind speeds. The observed small-scale structural variations can be broadly categorized as due to one of three sources: 1) static spatial structures apparently evolving from the ecliptic viewpoint due to solar rotation; 2) magnetic. reconnection driven flows (Antiochos et al. 2011); and 3) dynamic evolution such as turbulence. ***Solaris distinguishes between these sources*** because structures from rotation (i.e., source 1) can only be identified from a viewpoint parallel to the rotation axis. Moreover, ***Solaris resolves longitudinal structure of the streamer belt*** and associated solar wind flows, making it possible to determine the longitudinal expansion with radial height of magnetic flux tubes open to the solar wind. (See also Viall et al. (2022) WP submitted to this Decadal on how polar measurements helps solve Outstanding Problems in Solar Wind Physics.)

***Solaris' in-situ measurements*** of the solar wind's magnetic field, plasma composition, and kinetic properties—including radial speed variations will test 1) the extent to which magnetic field channels the fast and slow wind or whether this channeling arises from quasi-steady vs episodic phenomena indirectly linked to expansion (Wang and Sheeley 1990; Linker et al. 2011; Rappazzo et al. 2012); 2) provide additional mappings of magnetic topology from solar atmosphere to solar wind, helping to ***resolve inconsistencies between Parker spiral theory and Ulysses observations*** (*e.g.*, Smith *et al*., 1997, Fig. 6, left); and 3) determine whether the open flux in the solar wind remains nearly constant as predicted by the Fisk flux transport model (Fisk and Schwadron 2001, Fisk and Kasper 2020).

***Solaris has a unique view on large-scale dynamics.*** The impact of large-scale transient structures such as CMEs and corotating interatction regions (CIRs) on the global heliosphere and geospace is well documented (Miyoshi & Kataoka 2005; Schwenn 2006; Pulkkinen 2007). Conversely, the global corona and solar wind impacts the origins and evolution of these large-scale transients. This is obvious for CIRs, which originate at the interface of fast and slow wind streams. It is also true for CMEs, whose morphologies are known to change in the solar wind due to distortion (Howard & DeForest 2012), deflection and rotation (Isavnin et al. ***Solaris' unique*** out-of-the-ecliptic or ***"sunny-side-up" view*** (Fig. 5, right) observes ***CMEs and CIRs perpendicular to their direction of propagation***, yielding an unobstructed view on longitudinal trajectory, extent, and the dynamical evolution of these structures. Moreover, by combining *in-situ* measurements of high latitude interplanetary CMEs with EUV & white light coronagraphic views of their sources, ***Solaris sheds light on how global coronal structure vs. solar rotation impacts the dynamic evolution of CMEs*** as they move outward from the Sun.

## 3 Technical Approach

*Solaris* is a single spacecraft mission to fly over the Sun's poles to observe one of the last unexplored regions of the solar system. The *Solaris* technical approach and mission design is based on the results of a funded MIDEX Phase A mission study. *Solaris* mission described in this White Paper is an expanded/enhanced version of the MIDEX mission with the following differences;

1) the addition of 3 *in-situ* instruments (MAG, IES, FIPS), and 2) the extension of the design mission life from 5 to 10 years to enable three VGAs to circularize the orbit and extend mission operations to cover the solar cycle (solar min to max).

## 3.1 Instruments

*Solaris* includes the following focused set of remote sensing and *in-situ* instruments targeted to address ***Solaris'*** scientific questions that can only be answered from a polar perspective.

***Table 1****. Solaris Instrument Payload*

| Instrument | Solaris Measurements | Mass (kg) | Power (W) |
|---|---|---|---|
| **Remote Sensing:** | | | |
| Compact Doppler Magnetograph (CDM) | **Dopplergrams** with multi-month, high-latitude continuity<br>**Magnetograms** to quantify polar magnetic flux | 16.9 | 6.7 |
| EUV Imager (S-EUVI) | **EUV images** of coronal structure on disk & out to >3.0 solar radii | 11.4 | 13.2 |
| White Light Coronagraph (S-COR) | 360° out-of-the-ecliptic **WL coronal observations** from 2.5 to >15 solar radii, to overlap with EUV images & provide continuous coverage of longitudinal expansion from the low, through the middle to the high corona and solar wind | 10.1 | 4.0 |
| ***In-situ*:** | | | |
| Magnetometer (MAG) | *In-situ* vector **magnetic field** throughout *Solaris* orbit | 0.65 | 1.71 |
| Ion-Electron Spectrometer (IES) | **Solar wind** (electron, proton) speed, density, temperature | 1.04 | 1.85 |
| Fast Imaging Plasma Spectrometer (FIPS) | **Composition** & kinetic properties of solar wind heavy ions | 1.41 | 2.1 |

## 3.2 Spacecraft

*Solaris* is a 3-axis stabilized, Sun-pointed, solar-powered observatory, leveraging existing, high heritage, technology and subsystems. The primary drivers of the *Solaris* Observatory design unique to this mission are the environments at Jupiter, including reduced solar flux input for the solar arrays and the radiation environment. During the Jupiter flyby, beyond 5 AU from the Sun the observatory operates in a low-power, 3-axis controlled configuration.

## 3.3 Mission Design & Operations

The *Solaris* mission design uses a ballistic trajectory to achieve a direct injection to Jupiter for a JGA to escape the ecliptic plane, and multiple VGAs to circularize the orbit and reduce the orbital period to three years. The launch window for this mission design occurs every ~13 months. For the purposes of this White Paper, we are assuming a launch in 2030. After 2.5 months of commissioning and calibration, *Solaris* embarks on a 15-month Cruise to Jupiter. The JGA rotates the plane of the heliocentric orbit to an inclination of ≥70° to the solar equator. After the JGA and another 2.2-year Cruise, *Solaris* reaches its required heliographic solar latitudes of >55° for the remote sensing instruments, and performs a final instrument checkout and calibration. The remote sensing encounter phase begins over each pole when the spacecraft is >55 deg. heliographic latitude, and downlink occurs aperiodically during each pass to avoid artifacts in the helioseismology data. The *Solaris* ***operations concept*** consists of essentially ***two phases***: 1) ***Quiescent Cruise*** including JGA, beyond ~3 AU, and 2) ***Encounter Phase*** with two >100-day high latitude remote sensing observing periods, per orbit (Figure 2). The *in-situ* instruments remain powered on and collecting data during both encounter phase and during quiescent cruise. After commissioning, mission operations is assumed to follow a simple, repetitive flow with DSN contacts during cruise occurring every other week.

### 3.4 Mission Management

To enable rapid implementation of *Solaris* while PSP and SO are still operational, we recommend a PI-led mission paradigm based on the Planetary Discovery mission concept.

### 3.5 Technology Maturity

The majority of *Solaris* technology is already mature and high heritage, requiring no additional development. Almost all spacecraft and instrument subsystems, subassemblies, and components are matured to at least a Technology Readiness Level (TRL) of 6. The CDM TRL has been raised to 6, through the efforts of a SwRI Internal Research Program Grant.

### 3.6 Cost

The total *Solaris* mission life cycle cost (excluding launch) is ~$750M (FY22), including 35% reserves and ~$100M for instruments. Costs were derived starting with Solaris MIDEX Phase A costs, using historical data for project management, systems engineering, safety & mission assurance, and mission operations. Additionally, bottoms up cost modeling, including for *Solaris* instruments and spacecraft bus, was carried out in conjunction with the Mission Design Laboratory at Goddard Space Flight Center. It is noted that Phase E costs will be minimized with extended quiescent cruise phases and simple, encounter periods with only synoptic observations. Details on assumptions contained within the Solaris MIDEX CSR are available to NASEM upon request.

### 3.7 Community Impact

The Sun is not alone in generating magnetism; nearly all solar-system planets sustain some form of magnetic field (Jones *et al.,* 2011). The Earth's dynamo sustains a magnetic field that undergoes polarity reversals on much longer timescales and more irregular intervals than the Sun. Planets and stars possess two features in common that are central to the dynamo process: rotation and a convectively unstable, electrically conducting fluid. *Solaris* provides the first direct link between a magnetic dynamo and the convective flows that underpin it, yielding unique observational context for the broader range of stellar and planetary dynamos (Fig. 6).

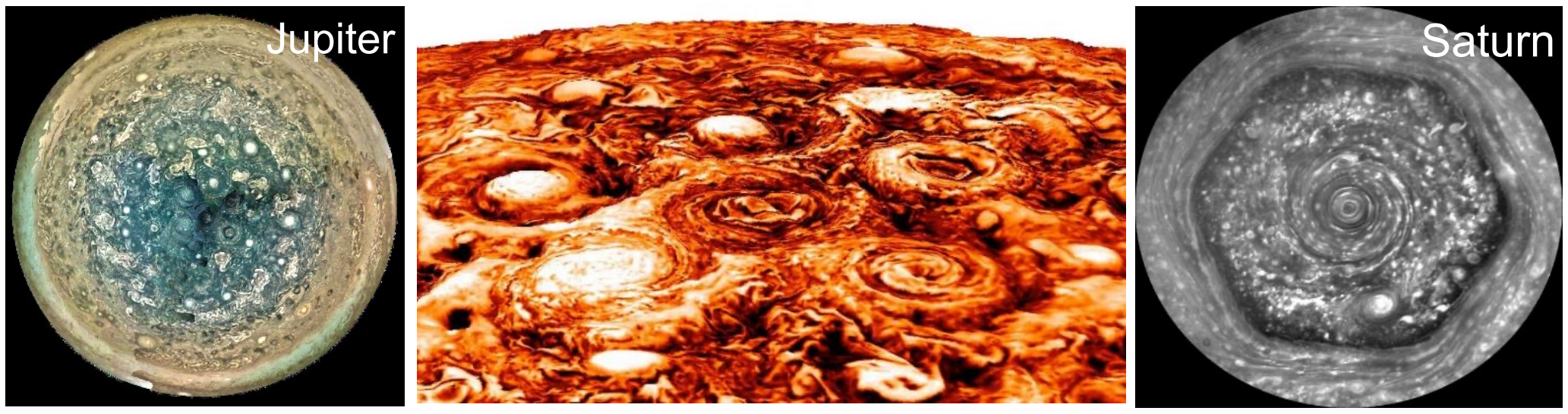

***Figure 6.*** *Just as our understanding of Jupiter and Saturn are being **transformed by polar observations** from Juno and Cassini, we expect our understanding of the Sun, the **solar dynamo** and **solar activity** will be **transformed** by the **polar observations from Solaris**.*

## 4 Summary

Enabled by a combination of new instrument technologies, new lower cost satellite buses and new high capacity, low $/kg-to-orbit launchers, *Solaris* is ready to fly now as a single spacecraft mission, or as the first spacecraft of a disaggregated constellation mission to achieve the highest priority solar science now, while key elements of the Heliophysics Systems Observatory (HSO) are still flying. ***We recommend Solaris to be considered as the first element of a new PI-led Heliophysics mission line (Kepko et al. 2022), analogous to the enormously successful Planetary Science Discovery-class mission line, with a Phase A-F cost cap of $800M (FY25), excluding launch vehicle.***

## References:

Antiochos, S. K., Mikić, Z., Titov, V. S., Lionello, R., Linker, J. A. (2011) A Model for the Sources of the Slow Solar Wind. *The Astrophysical Journal*, Volume 731, p.112.

Baker, D. N., A. Charo, and T. Zurbuchen. "Science for a technological society: The 2013–2022 Decadal Survey in Solar and Space Physics." *Space Weather* 11, no. 2 (2013): 50-51.

Baldner, C., et al. (2022). "Helioseismology of the Solar Poles", White Paper submitted to the 2024-2033 Solar & Space Physics Decadal Survey

Bale, S., et al. (2021). https://iopscience.iop.org/article/10.3847/1538-4357/ac2d8c

Charbonneau, P. (2020) Dynamo Models of the Solar Cycle. *Living Reviews in Solar Physics*, Volume 17, article id. 4. https://doi.org/10.1007/s41116-020-00025-6

DeForest, C. E., Howard, R. A., Velli, M., Viall, N., Vourlidas, A. (2018) The Highly Structured Outer Solar Corona. *The Astrophysical Journal*, Volume 862, p. 18.

Fargette, et al. (2021). https://iopscience.iop.org/article/10.3847/1538-4357/ac1112

Featherstone, N. A. & Hindman, B. W. (2016) The Emergence of Solar Supergranulation as a Natural Consequence of Rotationally Constrained Interior Convection. *The Astrophysical Journal Letters*, Volume 830, p. L15.

Featherstone, N.A., et al. (2022), "The Puzzling Structure of Solar Convection: Window into the Dynamo", White Paper submitted to the 2024-2033 Solar & Space Physics Decadal Survey

Fisk, L. A., and Schwadron, N. (2001) The Behavior of the Open Magnetic Field of the Sun. *The Astrophysical Journal,* Volume 560, p. 425.

Fisk, L. A. and Kasper J. C. (2020) The Global Circulation of the Open Magnetic Flux of the Sun, *The Astrophysical Journal*, Volume 894**,** idL4.

Gizon, L., Cameron, R. H., Majid Pourabdian, Zhi-Chao Liang, Fournier, D., Birch, A. C., Hanson, C. S. (2020). Meridional Flow in the Sun's Convection Zone is a Single Cell in Each Hemisphere. Science, Volume 368, p. 1469.

Howard, T. A. & DeForest, C. E. (2012) Inner Heliospheric Flux Rope Evolution via Imaging of Coronal Mass Ejections. *The Astrophysical Journal*, Volume 746, p. 64.

Jones, C.A. (2011) Planetary magnetic fields and fluid dynamos. *Annual Reviews of Fluid Mechanics*, Volume 43, p. 583.

Kepko, E. L., et al, (2022). "Heliophysics Needs a Mission Element between MIDEX and Flagship", White Paper submitted to the 2024-2033 Solar & Space Physics Decadal Survey

Linker, J. A., Lionello, R., Mikić, Z., Titov, V. S., Antiochos, S. K. (2011)The Evolution of Open Magnetic Flux Driven by Photospheric Dynamics. *The Astrophysical Journal*, Volume 731, p. 110.

Linker, J. A., Caplan, R. M., Downs, C., Riley, P., Mikic, Z., Lionello, R., et al. (2017). The Open Flux Problem. The Astrophysical Journal, 848(1), 70. https://doi.org/10.3847/1538-4357/aa8a70

Linker, J. A., Heinemann, S. G., Temmer, M., Owens, M. J., Caplan, R. M., Arge, C. N., et al. (2021). Coronal Hole Detection and Open Magnetic Flux. The Astrophysical Journal, 918(1), 21. https://doi.org/10.3847/1538-4357/ac090a

Linker, J.A., et al. (2022). "The Open Flux Problem: The Need for High Latitude Observations", White Paper submitted to the 2024-2033 Solar & Space Physics Decadal Survey

McComas, D.J., et al. (2008). "Weaker solar wind from the polar coronal holes and the whole Sun", GRL, 35, L18103, doi:10.1029/2008GL034896.

Miyoshi, Y., Kataoka, R. (2005) Ring Current Ions and Radiation Belt Electrons during Geomagnetic Storms Driven By Coronal Mass Ejections and Corotating InteractionRegions. *Geophysical Research Letters*, Volume 32, CiteID L21105.

Nandy, D. (2021). *Sol. Phys.*, **296**, 54, https://doi.org/10.1007/s11207-021-01797-2

Nandy, D., et al. (2022). "Exploring the Solar Poles: The Last Great Frontier of the Sun", White Paper submitted to the 2024-2033 Solar & Space Physics Decadal Survey.

Pulkkinen, Tuija (2007) Space Weather: Terrestrial Perspective. *Living Reviews in Solar Physics*, Volume 4, Article id. 1.

Rappazzo, A. F., Matthaeus, W. H., Ruffolo, D., Servidio, S., Velli, M. (2012) Interchange Reconnection in a Turbulent Corona. *The Astrophysical Journal Letters*, Volume 758, p. L14.

Schwenn, R. (2006) Solar Wind Sources and Their Variations Over the Solar Cycle. *Space Science Reviews*, Volume 124, p. 51.

Smith, E., Balogh, J. A., Burton, M. E., Forsyth, R., Lepping, R.P. (1997) Radial and Azimuthal Components of the Heliospheric Magnetic Field: Ulysses Observations, *Advances in Space Research*, Volume 20, p. 47.

Stakhiv, M., *et al* (2015). On the Origin of Mid-latitude Fast Wind: Challenging the Two-State Solar Wind Paradigm, *ApJ* **801** 100. doi:10.1088/0004-637X/801/2/100

Suess, S.T., S. Nerney (1999). Magnetic Fields and Solar Processes. The 9th European Meeting on Solar Physics, held 12-18 September, 1999, in Florence, Italy. Edited by A. Wilson. European Space Agency, ESA SP-448, 1999. ISBN: 92-9092-792-5., p.1101.

Upton, L., et al. (2022). "Improved Observations of the Sun's Polar Magnetic Fields", White Paper submitted to the 2024-2033 Solar & Space Physics Decadal Survey.

Upton, L., et al. (2022). "Meridional Flow: Polar Counter-cells", White Paper submitted to the 2024-2033 Solar & Space Physics Decadal Survey.

Wang, Y.-M., & Sheeley, N. R. . J. (1990). Solar wind speed and coronal flux-tube expansion. *The Astrophysical Journal*, *355*, 726. https://doi.org/10.1086/168805